# Evaluating Catchment Models as Multiple Working Hypotheses: on the Role of Error Metrics, Parameter Sampling, Model Structure, and Data Information Content

**Sina Khatami[1], Tim J. Peterson[1,2], Murray C. Peel[1], Andrew W. Western[1]**

[1] Department of Infrastructure Engineering, University of Melbourne, Parkville, Victoria, 3010, Australia

[2] Department of Civil Engineering, Monash University, Clayton, Victoria, Australia

Corresponding author: Sina Khatami (sina.khatami@unimelb.edu.au)

**Key points**

- KGEss is a more reliable metric than NSE and WIA, due to its mathematical structure.

- The choice of error metric — other things being equal — changes how model performance, parameter sampling sufficiency, and/or model hypotheses are measured.

- Relying on large samples of parameter space, without considering the model solution space, is a major source of uncertainty.

## Abstract

To evaluate models as hypotheses, we developed the method of *Flux Mapping* to construct a hypothesis space based on dominant runoff generating mechanisms. Acceptable model runs, defined as total simulated flow with similar (and minimal) model error, are mapped to the hypothesis space given their simulated runoff components. In each modeling case, the hypothesis space is the result of an interplay of factors: model structure and parameterization, chosen error metric, and data information content. The aim of this study is to disentangle the role of each factor in model evaluation. We used two model structures (SACRAMENTO and SIMHYD), two parameter sampling approaches (Latin Hypercube Sampling of the parameter space and guided-search of the solution space), three widely used error metrics (Nash-Sutcliffe Efficiency – NSE, Kling-Gupta Efficiency skill score – KGEss, and Willmott's refined Index of Agreement – WIA), and hydrological data from a large sample of Australian catchments. First, we characterized how the three error metrics behave under different error types and magnitudes independent of any modeling. We then conducted a series of controlled experiments to unpack the role of each factor in runoff generation hypotheses. We show that KGEss is a more reliable metric compared to NSE and WIA for model evaluation. We further demonstrate that only changing the error metric—while other factors remain constant—can change the model solution space and hence vary model performance, parameter sampling sufficiency, and/or the flux map. We show how unreliable error metrics and insufficient parameter sampling impair model-based inferences, particularly runoff generation hypotheses.

## 1 Introduction

The *summum bonum* (i.e. ultimate goal) of earth and environmental sciences, including hydrology, is to improve process understanding and prediction. Models are developed and improved by incorporating our understanding of real-world processes into them, and our understanding improves by modeling as a learning activity where models are treated as hypotheses of the real-world processes. Our understanding is ever-evolving, yet always remains incomplete and uncertain. While models are simplified representations of reality, they are most useful when used to challenge existing understanding (Oreskes et al., 1994). Due to this symbiotic and never-ending process of learning and modeling, developing frameworks for evaluating models as hypotheses under uncertainty is — and will always be — a research priority in hydrological sciences (Blöschl et al., 2019) and beyond.

Models can be evaluated from different standpoints. For instance, a *response space* (or *surface*) can be formed based on model parameters given some error metrics (Sorooshian & Gupta, 1983), or a *likelihood space* based on distributions of model parameters given some likelihood functions as a measure of model parameter uncertainty/sensitivity (Beven & Binley, 1992; Hornberger & Spear, 1981). Treating models as hypotheses, we developed a method to construct a *hypothesis space* based on equifinal model internal runoff fluxes that amount to the total simulated flow, called *Flux Mapping* (Khatami et al., 2019). The principle of equifinality implies that we should implement and evaluate models as multiple working hypotheses (MWH), which underpins the current paradigm of hydrological modeling (Beven, 2012; Buytaert & Beven, 2011; Clark et al., 2011a; Jehn et al., 2018; Krueger et al., 2010). A catchment model, including its internal fluxes and stores, is a simplified and approximate representation of catchment dynamics, averaged over spatio-temporal units. So, the internal runoff fluxes of hydrological models are indicative of catchment scale behavior for runoff generation, and hence provides a parsimonious way for testing and falsifying our knowledge of their corresponding catchment processes. In light of the above, the premise of this study is evaluating model runoff fluxes under uncertainty as MWH about catchment behavior/function namely runoff generation. It is a truism that model output is the result of the interplay between model structure and parameterization, data information content, and objective functions (or error metrics). The overall aim of this study is to unpack and demonstrate salient points of this interplay, which impact model-based inferences. We specifically address: how the error metric values change under different types or magnitudes of errors? What role does the error metric play in parameter sampling sufficiency? How error metric and/or parameter sampling influence model performance and process representation? To this end, we designed a series of controlled experiments to disentangle the role of each factor on the model output.

In the following sections we outline the dataset of 222 Australian catchments, runoff generation within the two hydrological models (section 2.2), three error metrics for model evaluation (section 2.3), and design of ensemble modeling experiments (section 2.4). A key contribution of this work is disentangling the role of error metrics, specifically their mathematical structure, in model evaluation and hypothesis formation. To this end, we conducted a one-factor-at-a-time sensitivity analysis on the mathematical structure of the three aforementioned error metrics (section 2.5), to demonstrate how each metric functions under different error types and magnitudes independent of any hydrological modeling (section 3.1). To the best of our knowledge a formal metric sensitivity analysis has not been done previously. Our results (section 3) show that some limitations in model evaluation and hypothesis testing are partly due to inherent characteristics of error metrics embedded in their mathematical structure — independent of model structure and parameterization, parameter sampling sufficiency, and forcing data. Such characteristics of error metrics may impede a reliable model evaluation, and thus give rise to misleading hypotheses. Finally, we discuss our findings including some of the limitations of this work that can be addressed in future studies (section 4).

## 2 Methods and experiment design

### 2.1 Study area and dataset

The study area is a subset of 222 unregulated catchments with relatively high-quality data over the period of record compiled by Fowler et al. (2020); the Australian edition of the Catchment Attributes and Meteorology for Large-sample Studies (CAMELS-AUS). In addition to the daily time series of observed streamflow of HRS catchments, the daily catchment average precipitation and daily Morton's areal potential evapotranspiration (APET) at the catchment centroid are also estimated. For further details on data preparation refer to Fowler et al. (2020). We limited our presented results (section 3) to a number of catchments that illustrate the impact of different modeling factors (i.e. model structure and parameterization, parameter sampling sufficiency, error metric, and forcing data) on flux maps (i.e. runoff generation hypotheses). A summary of catchment characteristics is presented in Table 1. Since it is not the aim of this study to evaluate the correspondence between catchment characteristics and model behavior, we do not further discuss catchment characteristics. Given that the aim of this study is to treat models as hypotheses (and not to calibrate models for predictions), we used the entire record of forcing data to calibrate/evaluate models.

**Table 1.** Summary of the study catchments used in modelling experiments and presented in the results section.

| Catchment No. | Corresponding figures | Catchment characteristics | | | | | | |
|---|---|---|---|---|---|---|---|---|
| | | Name | Location | Area (km²) | Mean annual precipitation (mm) | Mean annual streamflow (mm) | Mean annual APET (mm) | Annual runoff ratio |
| 1 | | Suggan Buggan River at Suggan Buggan | Victoria | 364.5 | 975.9 | 136.0 | 1088.5 | 0.14 |
| 2 | Figure 3 | Emu Creek at Emu Vale | Queensland | 153.8 | 996.2 | 99.2 | 1408.8 | 0.10 |
| 3 | | Currambene Creek at Falls Creek | New South Wales | 93.5 | 1075.1 | 202.5 | 1241.1 | 0.19 |
| 4 | Figure 4 | Wide Bay Creek at Kilkivan | Queensland | 352.3 | 945.0 | 147.3 | 1518.8 | 0.16 |
| 5 | Figure 5 | Kandanga Creek at Hygait | Queensland | 170.8 | 1135.2 | 278.0 | 1532.5 | 0.24 |
| 6 | Figure 6 | Normanby River at Battle Camp | Queensland | 2314 | 1533.6 | 364.4 | 1865.1 | 0.24 |
| 7 | Figure 7 | Elizabeth Creek at Mount Surprise | Queensland | 459.2 | 806.8 | 88.5 | 1641.9 | 0.11 |

## 2.2 Hydrological models: hypotheses of runoff generation

As hypotheses for runoff generation, a hydrological model may entail runoff generation mechanisms, whether at local or catchment scales, based on distinct catchment processes. In general, there are four main runoff generation mechanisms/sources: (1) Infiltration-excess overland flow, which occurs when rainfall intensity exceeds the soil infiltrability, also known as *Hortonian* overland flow (Horton, 1933). (2) Saturation-excess overland flow, also known as *Dunnian* overland flow (Dunne & Black, 1970), which occurs under saturated soil conditions, either due to direct rainfall (regardless of its intensity) on saturated soil, or due to the exfiltration (return flow) of a portion of interflow. (3) Subsurface stormflow, which is the rapid lateral movement/displacement of subsurface flow under saturated soil conditions (Hewlett & Hibbert, 1967). (4) Baseflow, which is the slow release of water from the catchment store.

For this study, we chose two conceptual hydrological models namely SIMHYD (Chiew et al., 2002; Peel et al., 2000) with 7 parameters, and SACRAMENTO (Burnash, 1995; Burnash et al., 1973) with 15 parameters. Despite their conceptual differences, these two are comparable process-based models for runoff generation, in that they simulate runoff through distinct runoff generating mechanisms. Total simulated flow in SIMHYD is the sum of three runoff fluxes representing different mechanisms of streamflow: (1) infiltration excess overland flow, (2) interflow and saturation excess overland flow, and (3) baseflow from a slow response reservoir. Details of SIMHYD and its runoff fluxes are explained in the literature (Chiew et al., 2002; Khatami et al., 2019; Peel et al., 2000). SACRAMENTO simulates runoff through five runoff fluxes: (1) runoff from permanently impervious areas (i.e. infiltration excess runoff), (2) direct runoff from additional impervious areas due to saturated conditions (a type of saturation excess runoff), (3) surface runoff when the Upper Zone Free Water storage is full (i.e. saturated conditions) and the precipitation intensity exceeds the rate of percolation and interflow, (4) interflow due to the lateral drainage of the Upper Zone Free Water storage, and (5) baseflow which is composed of primary and supplemental baseflow.

As Saffarpour et al. (2016) argued, catchment wetness drives both saturation excess overland flow (Western & Grayson, 1998; Western et al., 2005) and subsurface stormflow (Freer et al., 2002; Tromp van Meerveld & McDonnell, 2005). Infiltration-excess overland flow is an intensity-based mechanism, and baseflow is a slow (and often continuous) response, compared with event hydrograph timescales. Therefore, the runoff fluxes of these models can be classified into three groups or *modes* of model response, namely intensity-based, wetness-based, and slow response. Here we treat model output as a hypothesis indicating how runoff is simulated through these three modes of runoff generation for each modeling example. The flux map is a hypothesis space that summarizes an ensemble of acceptable/behavioral model runs based on their modes of model response (details in section 2.5).

### 2.3 Error metrics

We use three error metrics namely NSE (equation 1), skill score variant of KGE (KGEss, equation 2), and WIA (equation 3). Each metric quantifies some aspects of the (dis)similarity or distance between a *target* variable (e.g. observed streamflow time series, $O_i$ for $i = 1, \dots, n$ datapoints) and a *test* variable (e.g. modeled streamflow time series, $M_i$). NSE is based on least square errors, whereas WIA is built upon absolute errors (Willmott et al., 2012). Decomposing NSE, Murphy (1988) showed that NSE characterizes the distance between two variables (or time series) as an obfuscated function of their corresponding summary statistics: mean, standard deviation, and Pearson's linear correlation coefficient (CC). Refining the intrinsic redundancies within NSE, Gupta et al. (2009) developed KGE to systematically account for the three error terms of bias, variability, and correlation of two time series. In other words, KGE is inherently a multiple-criteria metric based on the Pareto set (or non-dominant solutions) approach (Gupta et al., 1998). Gupta et al. (2009) originally used standard deviation to account for the variability error. It was later substituted by the coefficient of variation to reduce the cross-correlation between bias and variability terms (Kling et al., 2012), which is the KGE variant that we used in this study (Equation 2.1).

$$NSE = 1 - \frac{\sum_{i=1}^{n}(M_i - O_i)^2}{\sum_{i=1}^{n}(O_i - \bar{O})^2} \quad ; -\infty \leq NSE \leq 1 \tag{Equation 1}$$

$$KGE_{ss} = 1 - \frac{1 - KGE}{\sqrt{2}} = ; -\infty \leq KGE_{ss} \leq 1 \tag{Equation 2}$$

$$KGE = 1 - \sqrt{\left(1 - \frac{\bar{M}}{\bar{O}}\right)^2 + \left(1 - \frac{M_{cv}}{O_{cv}}\right)^2 + (1 - CC)^2} \tag{Equation 2.1}$$

$$WIA = \begin{cases} 1 - \frac{\sum_{i=1}^{n}|M_i - O_i|}{2 \cdot \sum_{i=1}^{n}|O_i - \bar{O}|}, & when \ \sum_{i=1}^{n}|M_i - O_i| < 2 \cdot \sum_{i=1}^{n}|O_i - \bar{O}| \\ \frac{2 \cdot \sum_{i=1}^{n}|O_i - \bar{O}|}{\sum_{i=1}^{n}|M_i - O_i|} - 1, & when \ \sum_{i=1}^{n}|M_i - O_i| > 2 \cdot \sum_{i=1}^{n}|O_i - \bar{O}| \end{cases} \quad ; -1 \leq WIA \leq 1 \tag{Equation 3}$$

where $\overline{M}$ is the mean of the modeled series, and $M_{cv}$ and $O_{cv}$ are the coefficient of variation for the modeled and observed series respectively. All three are efficiency metrics, i.e. they assign a dimensionless scalar value to indicate the distance between the observed and modeled series. A perfect match would result in a metric value of 1, and as the modeled series diverge from the observed series the metric value decreases. NSE and WIA are inherently benchmarked against the mean of the observed series, $\overline{O}$. That is, the metric value is zero when the test (or modeled) series comprises of the overall mean of the target variable for every data point. Unlike NSE and WIA, KGE (both original and modified versions) is not benchmarked (Knoben et al., 2019). To benchmark KGE, here we developed the skill score version of KGE (KGEss, see Appendix A). Skill score is a common measure of the relative accuracy (or skill) of a forecast against a given reference/benchmark, e.g. NSE is essentially a skill score of mean squared error benchmarked against the observed mean (Murphy, 1988). KGE-based skill scores have been used previously for assessing the performance of hydrological models (Towner et al., 2019) and streamflow forecasts (Hirpa et al., 2018) benchmarked against some reference model/forecast. Here, we benchmarked KGE against observed mean to improve the comparability between the values of the metrics.

It should be mentioned that each metric characterizes some aspects of the distance between target and test variables, while no single metric can characterize all aspects (Khatami et al., 2019). We will further discuss this by cross comparing these three metrics in sections 3.1 and 4.1.

### 2.4 Experiment design

As shown in Figure 1, the experiment design has three main steps as follow:

**Step 1:** to setup the modelling experiments. To sample the parameter space, we generated two sets of Latin Hypercube Samples (LHS) of model parameter sets: 1 million LHS for SIMHYD, and 1.2 million for SACRAMENTO. These two sets of LHS parameter sets are used consistently for all modeling experiments, i.e. parameter sets do not vary across catchments and error metrics. Given the higher number of parameters in SACRAMENTO, we decided to use an additional 200,000 LHS parameter sets for SACRAMENTO. This is a subjective decision and does not guarantee sampling sufficiency, which varies by the choice of error metric, data information content, and model structure. The forcing data to the hydrological models are precipitation and evapotranspiration as explained in section 2.1, and the error metrics are NSE, KGEss, and WIA as explained in section 2.3.

**Step 2:** to run each hydrological model using two different parameterization approaches. (1) Random global search of the *parameter space* using the LHS parameter sets, resulting in an ensemble of model runs. (2) Guided global search of the *solution space* using Shuffled Complex Evolution (SCE, (Duan et al., 1992)) resulting in a single model run with the highest error metric value achievable. Due to inherent randomness in search routines like SCE, it is a common practice to repeat the search multiple times (Peterson & Fulton, 2019; Peterson & Western, 2014). Here, each modeling example was repeated 10 times for each error metric. The highest metric value among the 10 repeats (hereafter SCE-HMV) was chosen as the indicator of the guided search efficacy and a benchmark for the solution space, and the highest metric value of the model ensemble (hereafter Ensemble-HMV) as the indicator of the LHS effectiveness.

**Step 3:** to evaluate the model runs. As shown on Figure 1, model evaluation has three parts: (i) evaluating the sampling sufficiency, (ii) refining the LHS ensemble to define acceptable model runs, and (iii) flux mapping.

(i) Assessing the sample sufficiency by comparing Ensemble-HMV and SCE-HMV, i.e. comparing the best of the two worlds that accounted for both parameter space (based on the feasible range of parameter values) and solution space (based on the model performance given the model parametrization, error metric, and forcing data). We defined that a sampling is insufficient if for a given error metric | Ensemble-HMV – SCE-HMV | > 0.01. This is a relative test of sampling sufficiency where the sampling approach with the smaller indicator is certainly inadequate, while we cannot be certain about the adequacy of the other approach.

(ii) Refining the original LHS ensemble based on some criterion of model acceptability. For each error metric, the highest metric value (HMV = max{Ensemble-HMV, SCE-HMV}) achievable is an *upper benchmark* (Seibert et al., 2018) of the model performance (or solution space), regardless of the sampling strategy. This allows us to separate the influence of acceptability threshold from parameter sampling sufficiency on flux maps (i.e. model's runoff generation). The acceptability threshold is an arbitrary distance from the HMV for a given metric. For example, for the error metric KGEss we can apply a strict threshold of 0.03 (acceptability threshold = $HMV_{KGEss}$ − 0.03), or a more relaxed threshold of 0.10 (acceptability threshold = $HMV_{KGEss}$ − 0.10). A model run is defined acceptable if its corresponding metric value is above the acceptability threshold. While it is hard to objectively justify the choice of a threshold, we previously showed that the overall pattern of NSE-based flux maps is independent of the acceptability threshold (Khatami et al., 2019). Although it is clear that relaxing the threshold allows the acceptance of a larger number of model runs and relatively expands the flux map point cloud. We will further discuss the differences between these three error metrics and their impact on sampling sufficiency and model process-representation in section 3.2, using a variety of thresholds for different modeling examples.

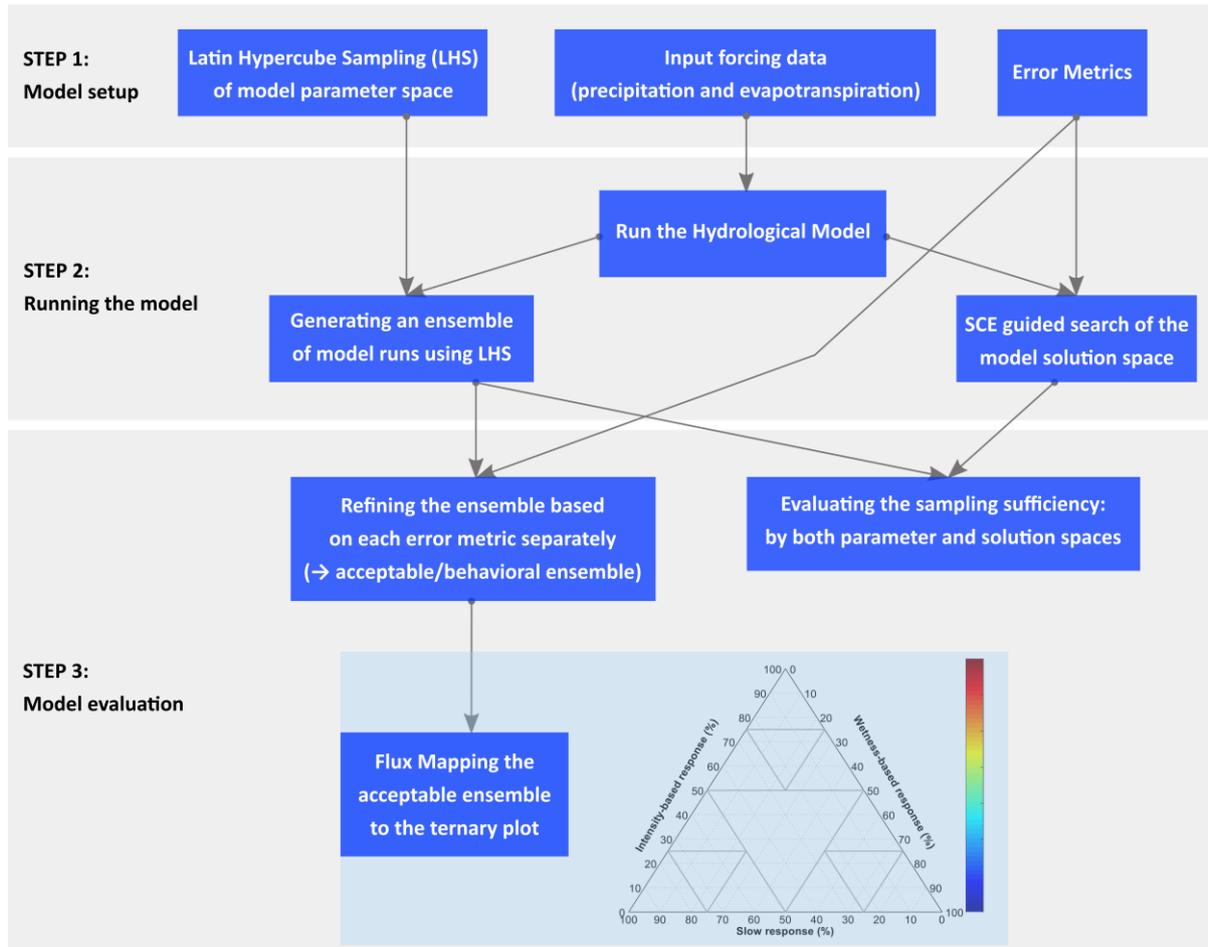

**Figure 1.** Schematic illustration of the modeling experiment design. The result of each experiment is to characterize the model response with a flux map.

(iii) Flux mapping the acceptable model runs to characterize how each model run simulates runoff generation (Khatami et al., 2019). Model parameters are often the only source of uncertainty that is accounted for, i.e. all sources of modeling uncertainty are implicitly lumped into the parameter uncertainty, although uncertainty sources such as model input (Kavetski et al., 2006; Khazaei & Hosseini, 2015; Moallemi et al., 2018; Papacharalampous et al., 2020a; Papacharalampous et al., 2020b; Vrugt et al., 2008), observed data (McMahon & Peel, 2019; Westerberg et al., 2016), and model structural uncertainty (Clark et al., 2015; Fenicia et al., 2011) can be accounted for more explicitly. Even when only parameter uncertainty is accounted for, flux mapping characterizes how uncertainty propagates from parameter space to flux space and hence the impact on model process-representation and MWH (Khatami et al., 2019). Each model run is represented as a point on the flux map (the ternary plot in Figure 1) based on the percentage of the volumetric contribution of each model runoff flux and color-coded by its performance (i.e. the error metric value). The upper value of the color bar is the Ensemble-HMV, and the lowest value is HMV – acceptability threshold. The flux map (triangle) is comprised of 4 smaller triangles, based on which the acceptable ensemble could be further classified as: (1) Slow response (or baseflow) dominated model response if more than 50% of the simulated runoff is produced by slow/baseflow response, i.e. the bigger bottom left triangle within the flux map. (2) Wetness dominated model response if more than 50% of the simulated runoff is produced by wetness-based runoff fluxes of the model, i.e. the bigger bottom right triangle within the flux map. (3) Intensity dominated response when more than 50% of the total simulated runoff is generated by intensity-based fluxes, i.e. the bigger upper triangle within the flux map. (4) No dominant mode when a model run is summarized into a point within the central triangle of the flux map. So, the flux map represents the relative dominance of different modes of model response that we defined in section 2.2.

It should be mentioned that as we used the SCE routine only for the global search of the parameter space (and not model calibration), its corresponding parameter set is not used in flux mapping.

### 2.5 Metric sensitivity

Here, we demonstrate how NSE, KGEss, and WIA function under three different error regimes namely *bias errors* ($e_B$), *variability errors* ($e_V$), and *correlation errors* ($e_C$). To this end, we took an arbitrary observed flow series, which includes multiple sequence of high and low flows, with 45 data points ($O_i, i = 1,2, ... ,45$), and conducted a one-factor-at-a-time sensitivity analysis (Pianosi et al., 2016) on each metric itself. In 20 steps ($k = 1,2, ... ,20$), we incrementally corrupted the observed series under each error type (see the example of step 1 in Figure S1). For bias errors, we corrupt the observed series to form a biased series (Series $B$), which is generated by adding a bias equal to 5% of the average of the original observed series, $\bar{O}$, at each step: $\overline{B^k} = (1 + k \cdot 0.05) \times \bar{O}$, while standard deviation and Pearson's linear CC with the original series were kept constant: $B_{std}^k = O_{std}$ and $corr_P(B^k, O) = 1$. In other words, increasing bias by 5% at each step under *ceteris paribus* (other factors held constant) assumption, i.e. standard deviation and CC unchanged. The residuals of series $B$ and $O$ represent bias errors, and the added bias at step 20 equals the mean of the original series ($e_B^{20} = \overline{B^{20}} - \bar{O} = \bar{O}$). For variability errors, we corrupt the observed series to form Series $V$, which is generated by increasing the standard deviation of the original series by 5% at each step: $V_{std}^k = (1 + k \cdot 0.05) \times O_{std}$, under *ceteris paribus* assumption: $\overline{V^k} = \bar{O}$ and $corr_P(V^k, O) = 1$. The residuals of series $V$ and $O$ represent variability errors, which is twice the standard deviation of the original series at step 20 ($V_{std}^{20} = 2 \cdot O_{std}$). For correlation errors, we corrupt the observed series to form Series $C$, which is generated by decreasing Pearson's linear CC between the original and corrupted series by 0.05 at each step: $corr_P(C^k, O) = 1 - k \times 0.05$, under *ceteris paribus* assumption: $\overline{C^k} = \bar{O}$ and $C_{std}^k = O_{std}$. The CC between the original series and the corrupted series at step 20 equals 0. The residuals of series $C$ and $O$ represent correlation errors, and $corr_P(C^{20}, O) = 0$. The original series and the three corrupted series are provided in the supporting information, Table S1.

## 3 Results

### 3.1 Metric Sensitivity: How do error metrics behave under different error regimes?

Comparing the corrupted series $B$, $V$, and $C$ with the original series $O$, Figures 2a-c show how the values of the three metrics degrade from their ideal value of 1 (step 0) under each error type. To further demonstrate the underlying mechanisms of the three error regimes, we also present the residuals for each error type and step (Figures 2d-f). For all error types, the original series remains uncorrupted at step 0, and hence the residuals for all data points (dark purple dots on Figures 2d-f) are 0, i.e. $B^0 = V^0 = C^0 = O$. Increasing the bias errors, enlarges the residuals homoscedastically (Figure 2d). That is, the magnitude of residuals increases while the variance of residuals remains constant; the zero slope of the linear lines highlighting the residuals at each step indicates this homoscedasticity. On the other hand, both variability and correlation errors generate heteroscedastic residuals (Figures 2e-f), but each exhibits a different type of heteroscedasticity. Variability errors lead to uniform (or linear) heteroscedasticity, indicated by a uniform increase in the slope of the highlighted lines in Figure 2e. Correlation errors, however, give rise to non-uniform (or non-linear) heteroscedasticity, indicated by a non-uniform expansion of the plain in which residuals lie (highlighted plains in Figure 2f). In short, bias errors are homoscedastic, variability errors are uniformly heteroscedastic, and correlation errors are non-uniformly heteroscedastic. It is worth mentioning that introducing correlation errors generates data points with negative values. While a negative flow is unrealistic, it does not matter for this particular sensitivity analysis.

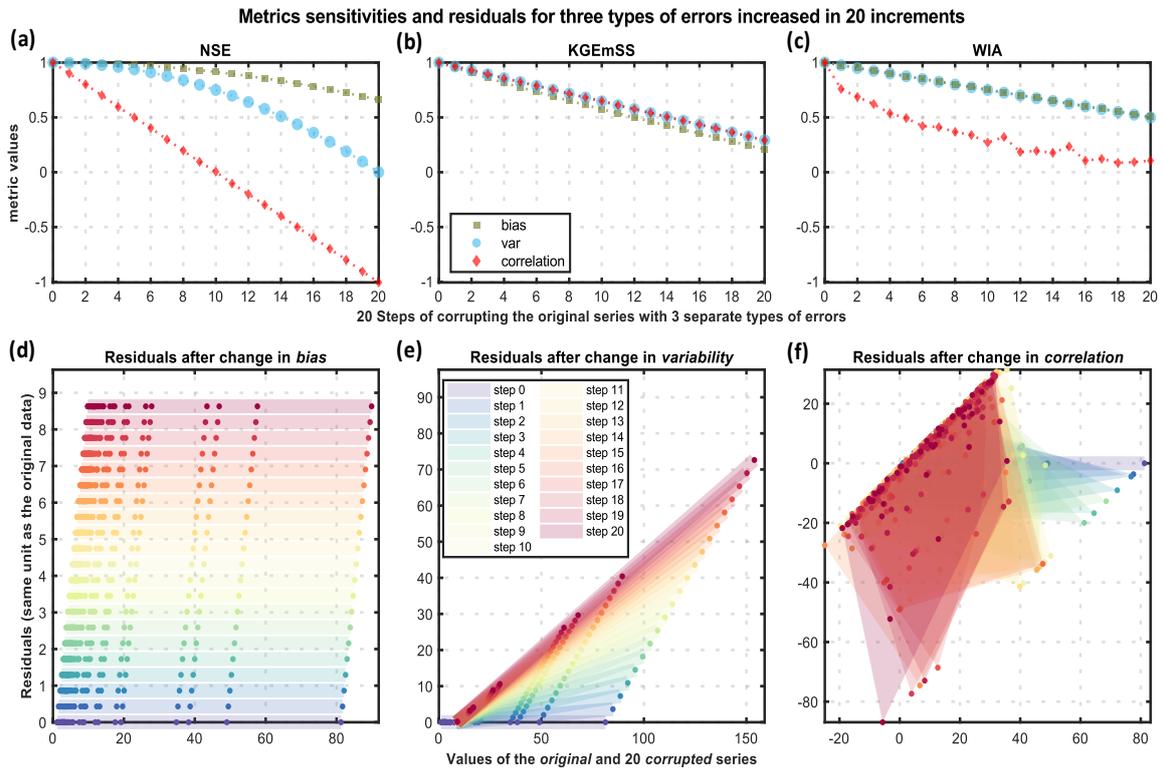

**Figure 2.** Sensitivity of efficiency metrics NSE, KGEss, and WIA in response to bias, variability, and correlation errors in 20 steps (a-c); the residuals of corrupted series for each error type and step (d-f). At step 0, corrupted series equals the original series ($B^0 = V^0 = C^0 = O$).

As shown in Figure 2a, NSE changes in a remarkably different way under the three error regimes, which arguably obscure the interpretability of NSE values. First, NSE exhibits varying degrees of sensitivity to different error regimes. At any given step, NSE is least sensitive to bias errors and most sensitive to correlation errors. The NSE's degradation line under bias errors (the line through green squares) has the smallest gradient of the three degradation lines. NSE values barely change for the first 5 steps, while KGEss and WIA values degrade more rapidly and linearly under bias errors. NSE is more sensitive to variability errors compared to bias errors, i.e. the degradation line of variability errors (the line through blue circles) has a steeper gradient. NSE is most sensitive to correlation errors as its degradation line under correlation errors (line through red diamonds) has the steepest slope between the three degradation lines. Due to this characteristic, for instance NSE = 0.80 can almost equally represent bias errors at step 16, variability errors at step 10, or correlation errors at step 3. In other words, a high NSE value does not equally represent the magnitude of the different type of errors. An NSE of 0.8 could contain a high bias, a medium variability error, or a small correlation error. This unequal sensitivity to different error types makes interpreting errors via NSE unreliable.

Second, NSE is less sensitive to bias and variability errors at higher NSE values (i.e. smaller error magnitudes) than lower values. This is due to the exponential decay of degradation lines of bias and variability errors, unlike the linear degradation line for correlation errors. In other words, although the magnitude of error is consistent across the error regimes and all 20 steps, NSE degrades inconsistently from one step to another for bias and variability errors (although consistently for correlation errors). For instance, a decrease in NSE values from $1.00 \rightarrow 0.90$ corresponds to larger bias or variability errors, than a decrease from $0.60 \rightarrow 0.50$. This characteristic obscures the interpretability and cross-comparison of NSE values across different ranges of itself. As we get closer to 1, it becomes harder to distinguish between models, whether comparing various model structures or parameter sets within a given model. Also, improving the performance of a given model, for example, from NSE: $0.50 \rightarrow 0.60$ is not comparable to NSE: $0.70 \rightarrow 0.80$. Due to this characteristic, a model can be accepted falsely (i.e. a false positive error) based on higher NSE values despite non-trivial bias or variability errors.

Third, comparing the three metrics, NSE is the least sensitive metric to bias errors and most sensitive to correlation errors at any given step (except for smaller correlation errors where WIA and NSE are not easily comparable due to irregular decay of WIA as shown on Figure 2c). This characteristic has important implications for cross comparing these metrics. While NSE may result in a high metric value despite relatively high bias errors, KGEss and WIA would yield lower values. On the other hand, NSE can generate lower values than KGEss and WIA under identical correlation errors. In other words, a model may be falsely rejected (i.e. a false negative error) because of lower NSE values due to NSE's over-sensitivity to correlation errors. While both KGEss and WIA consistently degrade under bias and variability errors, WIA degrades at a lower rate (compare the slopes of green squares and blue dots on Figures 2b-c). This implies that when comparing WIA and KGEss values under similar bias or variability errors, WIA will result in higher values due to its mathematical structure regardless of the actual performance of a model. The same comments apply to NSE and KGEss under correlation errors (compare the slopes of red diamonds in Figures 2a-b). So, using pre-determined metric values (despite recommendations such as NSE = 0.75 implying good model performance (Moriasi et al., 2007)) or cross-comparing metric values is not a reliable approach for evaluating model performance or improvement. We further demonstrate in section 3.2 that model performance and error metric value do not necessarily correspond.

Due to these three characteristics, achieving high NSE values does not necessarily imply smaller residuals, and hence does not imply a good model structure or performance (i.e. a false positive error). It could simply be due to the insensitivity of NSE to bias or variability errors at higher NSE values. On the other hand, a lower NSE value does not necessarily indicate a poor model structure or performance, as it can be due to the higher sensitivity of NSE to correlation errors (i.e. a false negative). In other words, NSE is an unreliable metric to evaluate model structure and characterize the model performance because of the inconsistent sensitivity of NSE to different error types and magnitudes, which is due to its mathematical structure and independent of the model structure or performance. NSE values are a result of complicated interactions between multiple bias, variability, and correlation terms inherent to the NSE function (see the NSE decomposition by Murphy (1988) and Gupta et al. (2009)). The problematic interaction between these components of NSE motivated the development of KGE, within which bias, variability, and correlation errors are separately and systematically accounted for.

Given its mathematical structure, KGEss functions consistently across all magnitudes (i.e. steps) of the three error types (Figure 2b). In other words, KGEss is equally sensitive to bias, variability, and correlation errors. The small difference between the degradation lines of bias errors and the other two errors is due to the variability term of KGEss being based on the coefficient of variation, which is a function of both standard deviation and bias. So, while standard deviation was kept constant under bias errors, the coefficient of variation (the variability term of KGEss) changes due to change in bias. Similar to KGEss, WIA functions consistently for different magnitudes of bias and variability errors (Figure 2c). But unlike KGEss, its degradation has an irregular (and somewhat exponential) decay under correlation errors. Although similar to KGEss, WIA degradation lines are linear across the steps, and WIA is less sensitive to both bias and variability errors than KGEss. In other words, even a small change in the decimals of WIA value indicates a relatively larger error, compared with the other metrics. This is due to WIA's mathematical structure being bounded at -1 for lower values, compared to the lower bound of NSE and KGEss being -∞. Such a narrow range of WIA values results in compact intervals and misleading interpretations if decimals are rounded. In this example, WIA = 0.75 may correspond to almost 50% increase in bias errors ($e_B = \sim 1.5 \times O_{mean}$), while KGEss = 0.75 can be due to about 25% increase in bias errors.

In summary, under the hypothetical conditions of this analysis: for similar bias errors, at each step NSE > WIA > KGEss; for smaller variability errors NSE > WIA > KGEss, and for larger variability errors WIA > KGEss> NSE; for correlation errors KGEss > WIA and KGEss > NSE, whereas for higher correlation errors KGEss > WIA > NSE, and for smaller correlation errors WIA and NSE are not easily comparable due to the irregular decay of WIA. Metric values for the degenerate cases (i.e. step 20) under each error regime are presented in Table 2. As shown, KGEss is the most consistent metric in terms of its sensitivity to different error regimes. While it is hard to generalize particularly beyond these three error types, it can be inferred that there would be a more controlled tradeoff between these error regimes under KGEss than the other metrics, which is due to its mathematical structure, and hence KGEss provides more reliable insights into model performance. That said, KGEss has its own limitations that we will discuss in section 4.1. Regardless of the limitations of error metrics, we argue that even a reliable error metric is not a sufficient condition for characterizing the model response.

**Table 2.** Metrics values for the degenerate cases (i.e. step 20) of each error type based on the original series mean ($O_{mean}$), standard deviation ($O_{std}$), and Pearson's correlation between the original and corrupted series at step 20 ($CC_p^{20}$).

| At step 20 | NSE | KGEss | WIA |
|---|---|---|---|
| **Bias errors = $O_{mean}$** | 0.66 | 0.21 | 0.50 |
| **Variability errors = $2 \times O_{std}$** | 0.00 | 0.30 | 0.50 |
| **Correlation errors: $CC_p^{20} = 0$** | -1.00 | 0.30 | 0.11 |

3.2 What determine the model response?

Here we demonstrate salient points of the interplay between model structure and parameterization, parameter sampling sufficiency, choice of error metric, and data information content. To this end, we conduct controlled experiments, i.e. varying one factor at a time while holding other factors constant (*ceteris paribus* assumption) to the extent possible, to disentangle the interplay of these factor. For each example, the model flux map is used to characterize the model response in terms of runoff generation. First (section 3.2.1), we examine the interplay of these factors for a single model SIMHYD, i.e. the model structure is unchanged. We then (section 4.2.2) examine the interplay of these factors considering both SIMHYD and SACRAMENTO, i.e. varying the model structure. For all examples the parameter sampling is controlled by using the same LHS parameter sets (1 M for SIMHYD and 1.2 M for SACRAMENTO) for all modelling experiments. For each catchment the data information content is controlled, i.e. the hydrological data (period, resolution, etc.) are the same. Details of each experiment are described accordingly.

3.2.1 Model response based on a single model structure

Figure 3 shows 9 different modeling examples: flux maps for 3 different catchments (each row) using SIMHYD with 3 error metrics (each column). For these 9 examples parameter sampling is considered sufficient as | Ensemble-HMV – SCE-HMV | ≤ 0.01. So, the HMV is within ±0.01 of the upper bound value of the color bar. For all examples the acceptability threshold is $HMV - 0.10$ (lower bound value of the color bar), and the model structure and parameterization is controlled i.e. SIMHYD with the same 1 M LHS parameter sets. For each row the data information content is also controlled (i.e. same catchment) and only the error metric varies, while for each column the error metric is controlled and the data information content across the three catchments varies. As shown on each row, for a given catchment and model parameterization, the choice of error metric can change the flux map in some examples (Figures 3a-c and 3d-f), while in some examples the choice of error metric is not as important (Figures 3h and 3i). On the other hand, the flux maps for two given catchments (#2 and #3) can be very different for some error metrics (NSE as in Figures 3d and 3g, and KGEss as in Figures 3e and 3h) and quite similar for another metric (WIA, as in Figures 3f and 3i). In other words, the interplay between the error metric and data information content for a given model structure and parameterization, can radically change the model response and hence the model's representation of runoff generation. So, when models are used to formulate hypotheses about catchment response, the hydrological (dis)similarity between two catchments can be radically changed by the choice of error metric — even under the same model structure and parameterization with sufficient parameter sampling.

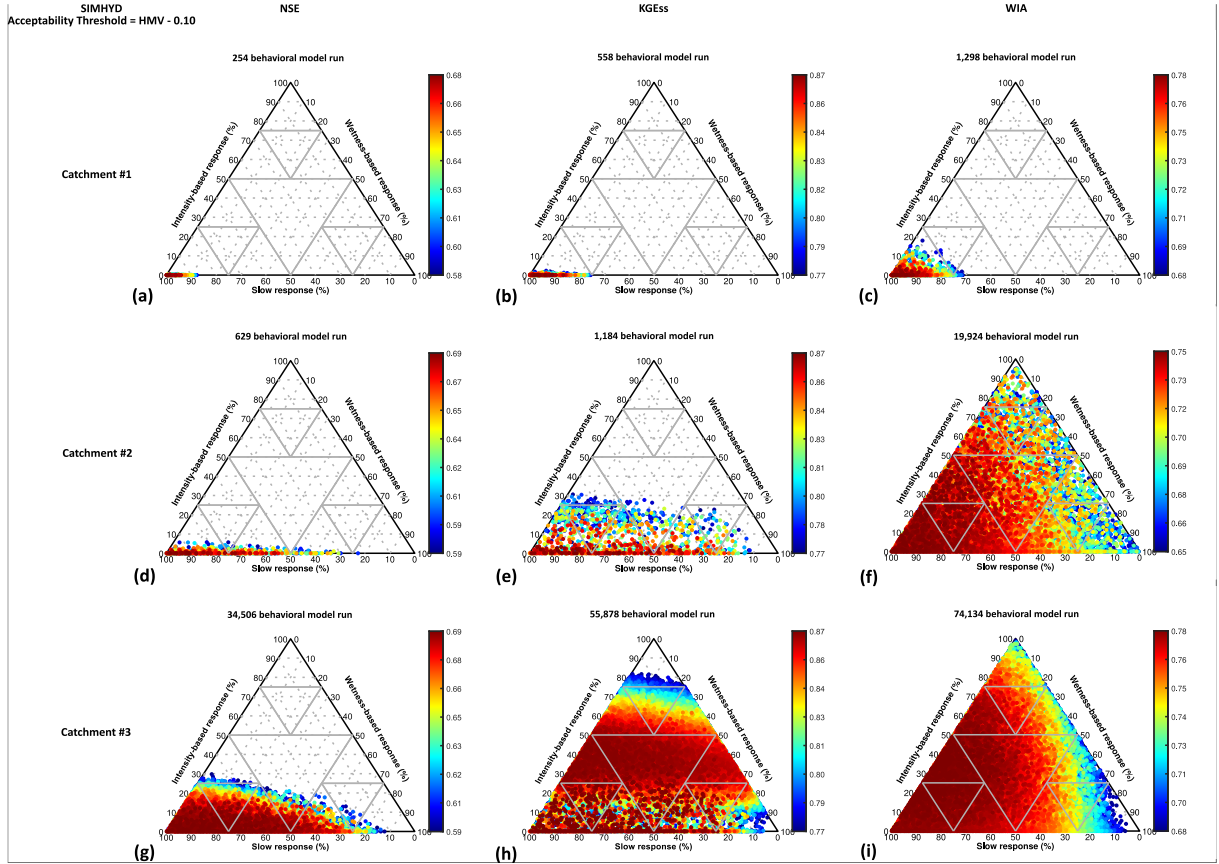

**Figure 3.** Model response (flux maps) of catchments #1-3 based on SIMHYD and the acceptability threshold of HMV – 0.10 for all error metrics. For all modeling examples parameter sampling is considered sufficient.

Given the behavior of error metrics at different intervals of their values (established in section 3.1), a given threshold would lead to a different number of acceptable runs under each error metric. Yet, as we discussed before (Khatami et al., 2019), the point cloud pattern of a flux map — and hence the model response — is not strictly dependent upon the number of acceptable model runs. The three examples of Figures 3f, 3h and 3i are space-filled flux maps with varying acceptable ensemble sizes from ~20,000 to ~74,000 model runs. A flux map can be space-filled with fewer acceptable model runs (Figure 3f, ~20,000 model runs), while another flux map can be constrained with more acceptable runs (Figure 3g, ~35,000 model runs).

In a different set of experiments (Figure 4) we gradually relax the acceptability threshold across the three metrics, under a *ceteris paribus* assumption (the catchment (#4), model structure and parameterization are unchanged). For each error metric (each column in Figure 4), the HMV is determined (HMV = max{Ensemble-HMV, SCE-HMV}), and the acceptability threshold relaxes in three steps from HMV – 0.03 to HMV – 0.06 and HMV – 0.09. As shown in Figure 4, the choice of error metric — even when other factors remain constant — can change the sampling sufficiency, which in turn can impact the flux map. For NSE (1st column in Figure 4), the 1 million LHS parameter sets are not sufficient as SCE-NSE – Ensemble-NSE ≈ 0.03; while for KGEss (2nd column in Figure 4) the SCE guided search is inadequate as Ensemble-KGEss – SCE-KGEss ≈ 0.02. So, for NSE the guided search and for KGEss the LHS was the better sampling approach for finding parameter sets with the highest metric values. The sampling sufficiency is considered sufficient for WIA (3[rd] column in Figure 4), which is at least partly due to compact intervals of WIA values as this metric is bounded (as explained in section 3.1). For the strict threshold (1[st] row in Figure 4), no model run is accepted under NSE (Figure 4a), whereas there are acceptable model runs under both KGEss and WIA (Figures 4b-c) but with different flux maps. So, given the choice of error metric, a set of LHS parameter sets not only may be (in)sufficient even for a model with only 7 parameters, but also can generate similar or distinct runoff generation hypotheses regardless of the sampling sufficiency. Given the degree of sampling insufficiency, all model runs may be rejected (i.e. no working hypotheses); not because of model structural inadequacy, but because of sampling insufficiency due to the choice of error metric (all other factors being held constant).

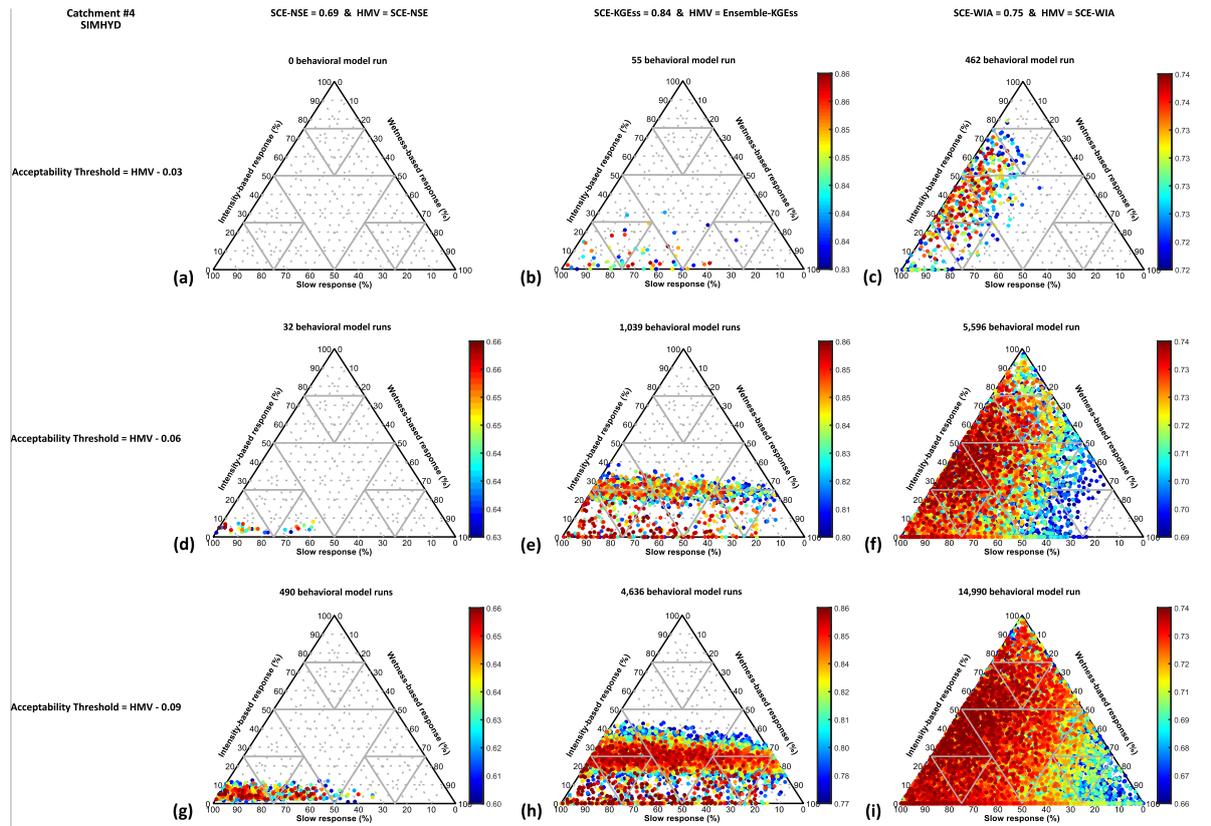

**Figure 4.** Model response (flux maps) of catchment #4 based on SIMHYD for the three error metrics (each column) with varying acceptability threshold (each row). Sampling sufficiency changes based on the choice of error metric: it is considered as sufficient for WIA, but insufficient for NSE and KGEss. For each error metric, HMV signals which parameter sampling (SCE or ensemble) was better i.e. found a parameter set with the highest metric value.

### 3.2.2 Model response based on multiple model structures

All the six modeling examples presented in Figure 5 are sufficiently sampled. The metric values for SIMHYD and SACRAMENTO are relatively similar under each error metric, yet the SIMHYD flux map is remarkably different from its corresponding SACRAMENTO flux map. For all error metrics, the SACRAMENTO intensity-based response is almost similarly constrained around 25% (Figures 5d-f). This is due to the fact that the intensity-based response in SACRAMENTO is determined as a fixed portion of the input rainfall by a constant parameter value, and hence there is not a wide range of variability for this flux. In SIMHYD, however, the runoff fluxes can interact widely. For SIMHYD each error metric gives rise to a different set of runoff generating hypotheses under the same model parameterization with sufficient parameter sampling (Figure 5a-c). For SACRAMENTO, on the other hand, the flux maps under the three error metrics are quite similar. For almost identical model performance under KGEss, SACRAMENTO gave rise to mostly wetness-dominated and slow response hypotheses, while SIMHYD resulted in a space-filled flux map i.e. any combination of model runoff fluxes is plausible to simulate the catchment response. So, while SIMHYD is a simpler model (smaller number of parameters, store, and fluxes), it exhibits a wider range of runoff generation hypotheses for catchment #5 even within a narrow range of (high) KGEss values.

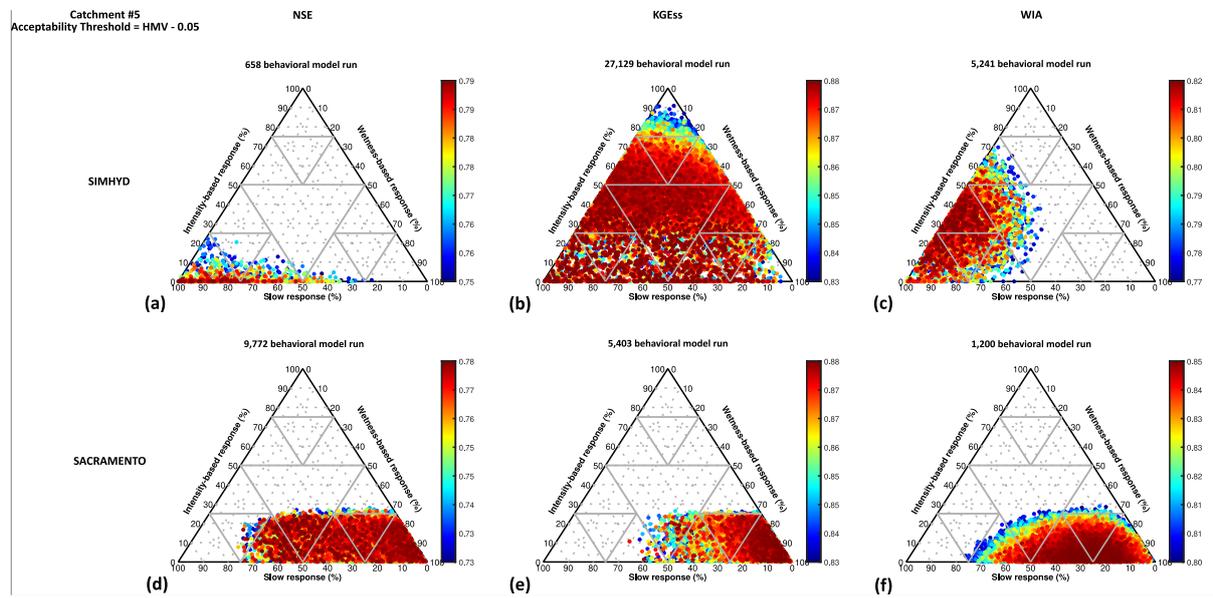

**Figure 5.** Model response (flux maps) of catchment #5 based on SIMHYD and SACRAMENTO, and the acceptability threshold of HMV − 0.05 for all error metrics. For all modeling examples parameter sampling is considered sufficient.

Although the 1.2 million SACRAMENTO LHS parameter sets were sufficient for catchment #5 under all error metrics (Figure 5d-f), they are insufficient for catchment #6 under NSE and KGEss (Figure 6d-e). This sampling insufficiency undermines both (A) model performance and (B) process representation. For catchment #6 and KGEss (Figures 6b and 6e): (A) the LHS ensemble misleadingly indicates a big difference between the performance of these two model structures (Ensemble-KGEss$^{SIMHYD}$ = 0.80 and Ensemble-KGEss$^{SACRAMENTO}$ = 0.69), against the SCE guided search indicating a relatively similar performance (SCE-KGEss$^{SIMHYD}$ = 0.81 and SCE-KGEss$^{SACRAMENTO}$ = 0.77). (B) Sampling insufficiency deflates the number of acceptable model runs under KGEss (only 4 even for a relaxed threshold, Figure 6e) resulting in a deficient flux map.

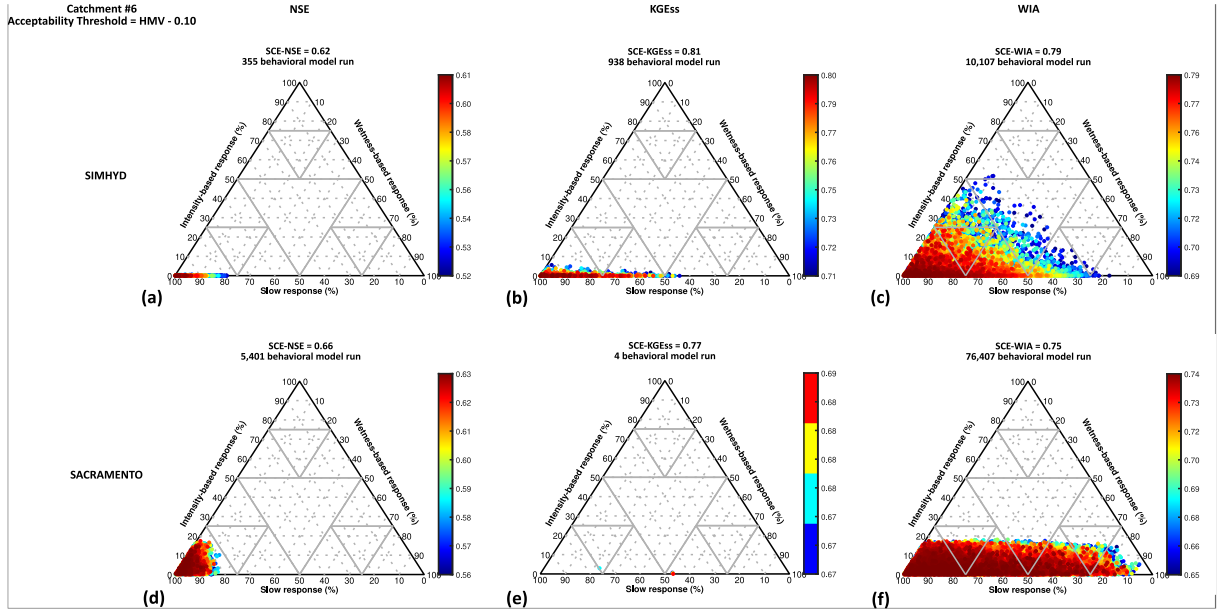

**Figure 6.** Model response (flux maps) of catchment #6 based on SIMHYD and SACRAMENTO, and the acceptability threshold of HMV – 0.10 for all error metrics. While for all SIMHYD examples parameter sampling is sufficient, it is not sufficient for SACRAMENTO under NSE and KGEss.

In catchment #6 and irrespective of sampling strategy, NSE suggests a better performance of SACRAMENTO in this catchment, while KGEss favors SIMHYD. That said, both models have equally high performance for catchment #7 under KGEss (KGEss = 0.92, Figures 7b and 7e) with sufficient parameter sampling. In a case like catchment #7, we can reliably compare the model structures and their processes representation (model flux maps) to formulate MWH about catchment response; because other factors are adequately checked i.e. equally high model performance and sufficient parameter sampling for a reliable error metric (KGEss) across all model structures. For catchment #7 and KGEss, the main distinction between these two models is that SIMHYD flux map indicates a catchment response with no significant intensity-based runoff generation, while SACRAMENTO suggests intensity-based response as large as 40% of the total flow. Such competing hypotheses can further be evaluated using additional data/knowledge about the catchment response.

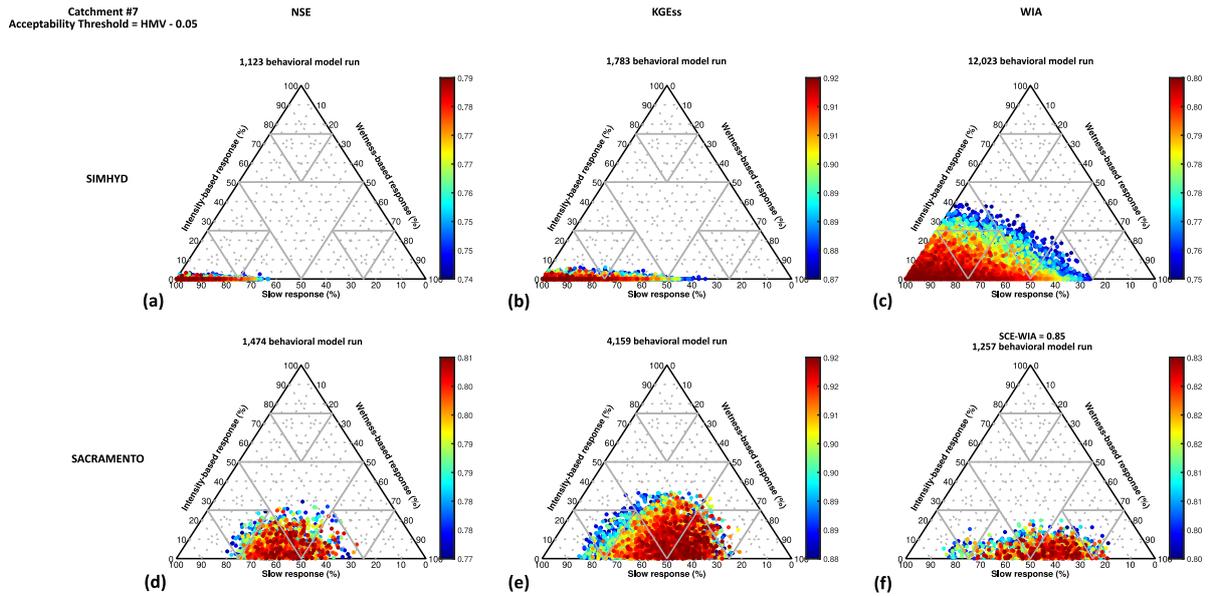

**Figure 7.** Model response (flux maps) of catchment #7 based on SIMHYD and SACRAMENTO, and the acceptability threshold of HMV – 0.05 for all error metrics. Parameter sampling is only insufficient under WIA and SACRAMENTO.

Based on analyzing 222 Australian catchments, we could not derive any systematic relationship between the error metric, number of acceptable model runs, sampling sufficiency, and size/type of the flux map point cloud across these two model structures (results not presented here). Examples of the range of interplay between these factors have been presented in Figures 3-7, from which we note some features. Firstly, Figures 3-4 shows that for a given error metric and under sufficient sampling, the flux map is independent of the HMV (i.e. model performance), acceptability threshold, or number of acceptable runs (also see Khatami et al., 2019). Secondly, the number of acceptable model runs is independent of the choice of error metric. Given that WIA intervals are very compact (bounded between +1 and -1), a certain range of WIA values can represent relatively larger errors and hence result in a higher number of acceptable model runs compared with NSE and KGEss; that said, this characteristic of WIA can be cancelled out by other factors and thus lead to a smaller number of acceptable model runs (e.g. compare Figures 5 and 7). The same comment applies to the impact of WIA on the size of the flux map point cloud (e.g. compare Figures 3i and 7f with comparable acceptable runs but different catchments and model structures). Thirdly, the number of acceptable runs is also not a function of the model structure, i.e. higher model dimensionality does not necessarily imply more flexibility in the model space and hence does not lead to more acceptable runs. With similar metric values, SIMHYD under KGEss (Figure 5b) has about five times more acceptable runs than its corresponding SACRAMENTO example (Figure 5e) (also compare Figures 5d-f and 6d-f for SACRAMENTO flux maps of two catchments under different acceptability thresholds, sampling sufficiency, number of acceptable model runs across the three metrics).

## 4 Discussion: evaluating catchment models as hypotheses under uncertainty

The model output and hence the generated MWH are the result of an interplay between model structure and parameterization, parameter sampling sufficiency, error metric, and data information content. As shown in section 3, this interplay is complex and unique to each case. That said, each factor can be controlled/improved to enhance model evaluation and hypotheses formulation. We further discuss a few points about each factor:

### 4.1 On the role of error metrics

*A robust error metric is a necessary condition for reliable model evaluation*. We conducted a one-at-a-time sensitivity analysis on the metrics NSE, KGEss, and WIA to characterize their behavior under well-defined error regimes, independent of any modeling. Willmott et al. (2015) opined that the interpretation of WIA is often more straightforward than NSE, and our sensitivity analysis is consistent with this: unlike NSE, WIA behaves consistently under bias and variability errors (Figures 2a and 2c). That said, we demonstrated that WIA's behavior hinders its interpretation in at least three ways: (a) WIA is more sensitive to correlation errors than bias and variability errors, (b) WIA's sensitivity to correlation errors is inconsistent across different intervals of WIA values, and (c) WIA intervals are very compact as it is bounded by $\pm 1$, hence WIA values degrade at a slower rate. We further discuss three major points about using error metrics for characterizing model performance:

(*i*) *NSE is a misleading error metric and the modeling community should abandon it*. There are perceptions about the meaning of NSE values, e.g. NSE $\geq 0.5$ indicates acceptable model performance (Davtalab et al., 2017; Moriasi et al., 2007) or acceptable parameter sets (Freer et al., 1996; Lane et al., 2019), the NSE = 0.6 as a threshold for acceptable model runs (Choi & Beven, 2007), NSE $\geq 0.75$ indicates good model performance (Moriasi et al., 2007), etc. Despite such widespread perceptions and based on a systematic sensitivity analysis of the NSE function, we demonstrated that NSE does not consistently represent different error types and magnitudes (Figure 2a and Table 2). As discussed, evaluating model performance based on higher NSE values may lead to false positives (e.g. accepting model runs and parameter sets despite large bias errors under-represented by higher NSE), or false negatives due to lower NSE values (e.g. rejecting models with small correlation errors exaggerated by NSE). Therefore, NSE is an unreliable metric to assess model prediction accuracy, benchmark model performance, or search the model solution space. From a process representation standpoint, given that NSE penalizes error regimes inconsistently, the solution space constructed based on NSE is unreliable due to its mathematical structure, even for a sufficient/representative parameter sample, regardless of data information content and the competence of the model structure. Shortcomings are inherent to models, and subjective decisions are inherent to various modeling decisions (Melsen et al., 2019; Moallemi et al., 2020a; Zare et al., 2020), including the choice of error metrics. That said, modelers can make better decisions. We believe that our study provides further evidence that NSE is inherently defective for model evaluation, and modelers and practitioners should instead use more reliable metrics such as KGEss, and ultimately aim to develop even better metrics.

(*ii*) *Cross-comparing error metrics is inherently problematic*. Error metrics behave differently under a given error type/magnitude due to differences in mathematical structure (Figures 2a-c and Table 2). So, it is inherently inappropriate to cross compare the values of different error metrics, unless their values are standardized to be compatible. For example, supposing that parameter set A gives NSE = 0.7, and parameter set B gives KGEss = 0.60 for a given model, can we infer that the model performs better using parameter set A? No. We can only cross compare A and B when they are both assessed with the same error metric. The same point also applies to cross comparison of various model structures using different error metrics.

*(iii) KGEss is a better metric than NSE and WIA, but it is not without its own flaws.* KGEss — unlike the other two metrics — responds consistently to at least three types of bias, variability, and correlation errors. So, KGEss values can be interpreted more judiciously, and we recommend using KGEss for single-metric evaluations. Furthermore, if parameter space is sufficiently sampled, the model solution space (i.e. acceptable parameter sets) and hypothesis space (e.g. runoff generation flux maps) derived based on KGEss are relatively more reliable, as they are at least independent of how KGEss behaves under different error types and magnitudes. However, the interaction between error terms within KGEss is not apparent in its final value. For instance, KGEss = 0.8 could equally be the result of various combinations of error terms e.g. with smaller or larger bias terms (a type of model-equifinality, see details in Khatami et al. (2019)). Yet, the tradeoff of the three error terms is relatively restrained/controlled under the mathematical structure of KGEss.

A major limitation of KGEss is that it does not explicitly account for the heteroscedasticity of model residuals, which is a general issue with almost all error metrics. Residual heteroscedasticity implies modeling inadequacy (i.e. potential to improve modeling setup), because there is information in the residuals (rather than residuals of random errors) that is not captured by the model structure and parameterization. This can be due to a combination of model structure and parameterization, error metrics, parameter sampling (in)sufficiency, and the fact that data themselves are not error free and their errors may propagate to the model outputs. While the issue of heteroscedasticity is long recognized (Sorooshian & Dracup, 1980), it is not explicitly accounted for in KGE nor WIA (or other metrics based on absolute error (Legates & McCabe, 1999)). Despite numerous reviews and comparisons of error metrics (Bennett et al., 2013; Crochemore et al., 2015; Gueymard, 2014; Krause et al., 2005; Moriasi et al., 2007), it is not clear what role the mathematical structure of error metrics particularly play in giving rise to heteroscedastic residuals. Two general approaches to address residual heteroscedasticity have been studied. (i) To indirectly account for heteroscedasticity by transforming flow series using transformations (McInerney et al., 2017) such as Box-Cox (Box & Cox, 1964; Yeo & Johnson, 2000), inverse function (Pushpalatha et al., 2012), or $n^{th}$ root functions (Chiew et al., 1993; Chiew et al., 1995), to put more emphasis on low flows and hence harness the heteroscedasticity of model residuals. While inverse function offers some improvements, particularly better results than logarithmic transformations, it has its own limitations (e.g. when flows become close to zero) for the estimation of the water balance, physical interpretation of error terms, and model calibration (Santos et al., 2018). (ii) There are also approaches to directly account for heteroscedasticity, which also have their own limitations. For example, Evin et al. (2014) proposed postprocessing model parameters for heteroscedasticity and autocorrelation but their approach works poorly in ephemeral catchments.

Given the above, there is room to further improve KGEss by developing a new error term to account for residuals heteroscedasticity or develop new error metrics, which is an important theoretical quest with significant practical implications for practitioners. In doing so, a few points should be considered:
- Redundant error terms should not be embedded in an error metric.
- Error metric should function consistently across different error types/magnitudes.
- Error metric should behave consistently across different periods of high and low flows.
- There is no ultimate metric, no matter how elegant a metric would be, it can only characterize certain (and not all) aspects of model-observation (dis)similarity. Therefore, it is essential to only use/interpret metrics that are fit for purpose.

### 4.2 On the role of model structure and process representation

In addition to error metric, model structure also influence the runoff generation hypotheses. For instance, as shown in Figures 5d-f, the intensity-based response of SACRAMENTO is similar across the three error metrics; while SIMHYD (Figures 5a-c) results in distinct flux maps and varying degrees of intensity-based response for each error metric. Because the partitioning of input rainfall into intensity-based runoff flux is determined by a constant model parameter in SACRAMENTO. That is, the SACRAMENTO model structure constrains the intensity-based runoff generation in these cases. That said, Figures 7e-f show that SACRAMENTO results in different flux maps, with almost twice as much intensity-based response under KGEss than WIA, in a different catchment. Whereas, SIMHYD allows for four times as much intensity-based response under WIA than KGEss, in the same catchment. So, the model structure plays a role in how model fluxes, and hence hypothesis of catchment processes, are allowed to behave for a given catchment and error metric. Undoubtedly, the representation of runoff generation mechanisms in these hydrological models are simplifications of real-world processes. Runoff generation varies in time (e.g. due to seasonality or land-use changes) and space (due to catchment heterogeneity), and often a mix of these processes causes runoff (Saffarpour et al., 2016). Particularly as we are using lumped daily models treating a catchment as a single spatial unit, heterogeneity and sub-daily variations of these processes are overlooked and aggregated into daily catchment averages. Despite such simplifications and other sources/types of modeling uncertainties, a conceptual model and its internal dynamics can still be indicative of different (dominant) catchment processes (Dunn et al., 2008; Guo et al., 2017; Lerat et al., 2012). Given that processes such as runoff generation are incorporated into conceptual models at least partly with the aim of improving realism, thus these internal components should be evaluated in addition to the final model output.

Use of process-based models for evaluating runoff generation mechanisms has been previously studied. For example, Grayson et al. (1992) compared the representation of different runoff generation mechanisms in a process-based model across a few Australian and north American catchments. Buchanan et al. (2018) characterized the predominance of infiltration-excess and saturation-excess runoff across the contiguous United States. With flux mapping, we formalize a hypothesis space based on different modes of model runoff fluxes (Khatami et al., 2019), that is useful for formulating and comparing MWH for runoff generation (catchment response) across different catchments, periods, and model structures. Treating models as hypotheses, modeling would be a learning activity to formulate alternative/competing hypotheses. Testing hypotheses against catchment behavior and attributes using field data (Clark et al., 2011b; Seibert & McDonnell, 2002; Winsemius et al., 2009) is the avenue towards evaluating the plausibility of these hypotheses, and to further improving model realism (Gharari et al., 2014; Hrachowitz et al., 2014; Wagener, 2003).

### 4.3 On the role of parameter sampling sufficiency

*Sufficient parameter sampling is a necessary condition for reliable evaluation of models as MWH*. Sampling insufficiency undermines both model performance and process representation, as demonstrated in the results (Figures 4, 6 and 7). A representative sample of the parameter space can be achieved either by guided search routines and/or large random samples. While we acknowledge that various methods have been developed to sample the parameters more effectively and efficiently (Asadzadeh & Tolson, 2013; Sheikholeslami & Razavi, 2017; Tolson & Shoemaker, 2007; Vrugt & Beven, 2018), we adopted two of the most widely used sampling strategies in hydrological modeling: large LHS to sample the parameter space and SCE to benchmark the solution space. We compared these two strategies against one another in each modeling case, i.e. compare { Ensemble-HMV, SCE-HMV }, as a test of relative sampling sufficiency.

An overview of our results across all 222 catchments show that large samples of parameter space were better only in 4% (or less) of cases (compare row 1 and 2 of Table 3), than the SCE search. This implies that it is a better approach to search the model solution space to either sample behavioral/acceptable parameter sets or benchmark the model performance. A geometry-based strategy like LHS aims to sample different regions of the parameter space more evenly than a random sample, yet LHS samples may even fail to be geometrically representative due to their inherent randomness (Goel et al., 2008), let alone sufficient for the model solution space (Tolson & Shoemaker, 2008). Relying on large samples of the parameter space, without considering the model solution space, is a major source of uncertainty for model evaluation and hypothesis formulation. Particularly, for higher model dimensionality (SACRAMENTO), the risk of relying only on large samples of the parameter space increases (the percentage of equal cases drops, e.g. from 52% to 34% for KGEss, Table 3). It is worth mentioning that in addition to model performance, WIA also obscures the evaluation of sampling sufficiency due to its compact intervals.

**Table 3.** The percentage of catchment models (out of 222 catchments) that were sufficiently sampled with a given sampling method relative to the other one. The criteria for relative sampling superiority is Ensemble-HMV – SCE-HMV > 0.01.

| Sampling strategy | SIMHYD | | | SACRAMENTO | | |
|---|---|---|---|---|---|---|
| | NSE | KGEss | WIA | NSE | KGEss | WIA |
| LHS ensemble of parameter space | 4% | 4% | 0% | 3% | 4% | 1% |
| SCE search of solution space | 62% | 44% | 13% | 74% | 62% | 49% |
| Both are equal (by a 0.01 margin) | 34% | 52% | 87% | 23% | 34% | 50% |

Inadequate sampling can lead to missing some plausible model runs, under-utilizing the model structure, and hence under-representation of MWH (e.g. Figures 4a, 4b, and 6e). This is important in large-sample studies as a particular ensemble of parameter sets, regardless of the sampling strategy, may be insufficient in some modeling cases; thus impacting the conclusions based on modelling results. It is also necessary to jointly evaluate the sampling sufficiency on both parameter and solution spaces for diagnostic evaluation of model failure in hypothesis testing and rejection based on models.

For instance, Hollaway et al. (2018) recently reported that given some limits of acceptability, no acceptable model run was found to simulate phosphorus load within a uniform random sample of 5 million sets for the SWAT model (based on 39 parameters). They concluded that the SWAT model structure is to be rejected as not fit-for-purpose. They primarily focused on the role of data information content, i.e. uncertainty in the calibration data, within the limits of acceptability approach. While the role of data uncertainty is undeniably crucial in model evaluation, they did not consider the role of parameter sampling sufficiency: (1) Is 5 million random parameter sets sufficient, just by the virtue of sample size, for sampling such a high dimensional parameter space? (2) Is *the* sampled set sufficient for the model solution space? It is therefore an open question whether or not a more adequate parameter sample would have avoided the model rejection and yielded some MWH in that study. One solution is to combine the best of the two worlds: to increase the LHS size sequentially, e.g. using Progressive LHS method (Sheikholeslami & Razavi, 2017), while comparing each sequence against a solution space benchmark.

4.4 On the limitations of this study and future directions

We acknowledge that in our sensitivity experiment (section 3.1) we introduced idealized errors, while in real-world cases errors could be more complex in nature. Streamflow data are uncertain (McMahon & Peel, 2019; McMillan et al., 2018; Westerberg et al., 2011) and may encompass different epistemic errors and disinformative periods (Beven et al., 2011; Beven & Westerberg, 2011), with complex interactions with each other and other factors involved in model behavior. That said, here we performed sensitivity analysis under ideal conditions to understand the function of each error metric independent of the quality of the data and the model structure. It would also be interesting to further understand the function of error metrics under common errors in hydrological residuals such as autocorrelation and heteroscedasticity errors.

We used the overall mean of the observed streamflow as the benchmark inherent in the error metrics, while it is a minimal benchmark (Schaefli & Gupta, 2007). We also did not differentiate between different periods in the data in terms of their information content or quality, nor consider the temporal dynamics of runoff generation. Future studies could look further into the dynamics of runoff generation across different seasons or multi-year periods with different characteristics. It would also be interesting to further study the correspondence between flux maps, i.e. dominant modes of model response, and catchment characteristics and attributes to further evaluate the plausibility of flux maps.

Here we evaluated catchment models as hypotheses based on three distinct modes of runoff generation embedded in model structures. Other internal components of process-based models such as evapotranspiration and soil moisture could also be evaluated. Characterizing and evaluating the internal model fluxes provides an avenue to evaluate model process-representation, diagnose model structural shortcomings, and ultimately improve process-based models.

We defined a sampling as insufficient if| Ensemble-HMV – SCE-HMV | > 0.01, i.e. based on the value of error metrics. While this can be seen as a test for sampling insufficiency, we emphasized that we cannot be certain about the adequacy of a sample based on this test. We chose the SCE guided search as it is widely used in Earth and environmental modeling. There are other methods that are shown to be more effective and efficient (Arsenault et al., 2013). While we certainly agree to embrace sampling efficiency (Tolson & Shoemaker, 2008; Vrugt & Beven, 2018), we further argue for embracing the uniqueness of the model response (and MWH), particularly in studies with large samples of catchments, models, and objective functions. Therefore, no matter how robust a search algorithm works under different numerical experiments, the parameter sampling sufficiency should also be evaluated for each modeling case given the choice of error metric and forcing data.

## 5 Conclusion

Here we demonstrated that model response is the result of a complex interplay between factors of model structure and parameterization, parameter sampling sufficiency, choice of error metric, and data information content. This interplay is unique to the underlying assumptions and conditions of each modeling case, and variations in each factor can remarkably change the model response. We argued that a hypothesis space can be constructed based on model internal (runoff generating) fluxes, that could be used to characterize and compare process-representation of different models under different assumptions. We demonstrated that deficient error metrics and insufficient parameter sampling undermine both model performance and process representation (model-based hypotheses). Conducting sensitivity analysis on the mathematical structure of three widely used error metrics, we demonstrated that KGEss is a more reliable metric than NSE and WIA, even though KGEss has its own limitations. Furthermore, relying on large Latin Hypercube samples of the parameter space, without considering the model solution space, is a major source of uncertainty. It is ultimately our goal to advance theoretical frameworks for process-based evaluation of models as hypotheses to better understand and model human-natural systems under uncertainty and non-stationarity (Khazaei et al., 2019; Lu et al., 2018; Moallemi et al., 2020b; Westerberg et al., 2017).

## Acknowledgements

The authors gratefully acknowledge the support of the University of Melbourne and Australian Government in carrying out this research; Sina Khatami is supported by Melbourne International Research and Fee Remission Scholarships (MIRS and MIFRS), Murray Peel the recipient of an Australian Research Council Future Fellowship (FT120100130), and Tim Peterson jointly funded by Australian Research Council Linkage Project LP130100958, Bureau of Meteorology (Australia), Department of Environment, Land, Water and Planning (Vic., Australia), Department of Economic Development, Jobs, Transport and Resources (Vic., Australia) and Power and Water Corporation (N.T., Australia).

## Data availability

Data for streamflow, rainfall data, and potential evapotranspiration are all available at https://doi.pangaea.de/10.1594/PANGAEA.921850.

## Appendix A: deriving the equation for KGE skill score (KGEss)

Skill score refers to the relative accuracy of model predictions (or forecasts) for a particular measure of accuracy (A) given a reference value ($A_{ref}$) and perfect value ($A_{perf}$), and is measured as:

$$skill\ score = \frac{A - A_{ref}}{A_{pref} - A_{ref}}$$

For A = KGE with $KGE_{pref} = 1$ and benchmarked against observed mean $A_{ref} = KGE(\overline{O}) = 1-\sqrt{2}$, the KGE skill score (KGEss) derives as below:

$$KGEss = \frac{KGE - (1 - \sqrt{2})}{1 - (1 - \sqrt{2})} = \frac{KGE - 1 + \sqrt{2}}{\sqrt{2}} = 1 - \frac{1 - KGE}{\sqrt{2}}$$